# Origin of superionic state in Earth's inner core

Ina Park[1], Yu He[2,3], Ho-kwang Mao[2], Ji Hoon Shim[1,4]*, Duck Young Kim[2]*

**Affiliations:**

[1]Department of Chemistry, Pohang *University of Science and Technology, Pohang 37673, Republic of Korea.*

[2]*Center for High Pressure Science and Technology Advanced Research (HPSTAR), Shanghai 201203, China.*

[3]*Key Laboratory of High-Temperature and High-Pressure Study of the Earth's Interior, Institute of Geochemistry, Chinese Academy of Sciences, Guiyang 550081, China.*

[4] *Department of Physics, Pohang University of Science and Technology, Pohang 37673, Republic of Korea.*

*Corresponding author. Email: Ji Hoon Shim jhshim@postech.ac.kr, Duck Young Kim duckyoung.kim@hpstar.ac.cn

**Abstract:** Earth's inner core (IC) serves as a reservoir for volatile elements, which significantly affects its behavior and properties. Recent studies suggest that superionicity can be observed in ice and iron hydrides under high-pressure and temperature conditions, providing an alternative understanding of the planet's interior. In this study, we demonstrated that electride formation drives the superionic state in iron hydride under IC pressure conditions. The electride stabilizes the iron lattice and provides a pathway for volatile diffusion. The coupling between lattice stability and superionicity is triggered near 100 GPa and enhanced at higher pressures. The electride-driven superionicity can also be generalized for volatiles in other rocky planetary cores. These findings provide new insights into the mechanisms of core formation and evolution of rocky planets.

Electride is a material with electrons occupying the interstitial sites and playing a role as anions. These electrons are localized at so-called non-nuclear attractor (NNA) sites, the state with high electron-localizability (1-8). NNAs generate novel properties due to their loosely binding characteristics, so potential applications such as catalysts, reducing agents, electron emitters, battery anodes, and even superconductors have been proposed and discovered (9-18). To induce NNA, high pressure conditions can be one of the intrinsic platforms because valence electrons gain enough kinetic energy to escape attracting potential well of the atomic nuclei and occupy vacant space in solids. This mechanism was theoretically modeled (7) and computationally verified for real metals, for example, by predicting open-structured aluminum (19). Furthermore, a formation of open-structured magnesium was recently observed at tera-pascal pressure with the presence of NNA experimentally (20).

On the other hand, the chemical and physical role of NNA on structure and material properties was found to be intricate and prominent when it meets hydrogen atoms. Since the relationship between a hydrogen atom and NNA was proposed by Savin *et al*. (21), many studies have shown indirect clues about the relations. It was found that the structural, electronic, and dynamic properties of atoms were governed by the NNAs of the parent compound (22-25). The study, particularly regarding the relationship between the electride and superionic state, implies that the NNA may affect not only the electronic property but also the energetics and kinetics of the material. However, to date, few studies have discovered the coexistence of both states (16, 25), and the relationship between the two states has been even more rarely discussed.

In this work, we found that the high pressure superionic behavior of H anions in $hcp$-FeH$_x$ (x < 1) (26) can be a comprehensive example to study both superionicity and electride phenomena and their entanglement. Iron-light-element alloy, $hcp$-Fe(H, C, or O)$_x$ (x < 1), is one of the Earth's inner core (IC) material candidates, and it becomes superionic under a high pressure condition of around 300 GPa. A superionic state is widely reported to exist in planetary interiors with significant influence on the properties of interior materials (26-29). We found that it also becomes an electride at high pressure conditions of about 100 GPa and that the doped hydrogen ions were governed by the NNAs of the parent compound *hcp*-Fe. By examining the stabilization effect of NNA to maintain the metal lattice, we propose that NNA can play a role in offering a pathway to superionicity.

**I. Possible entanglement of electride and superionic state**

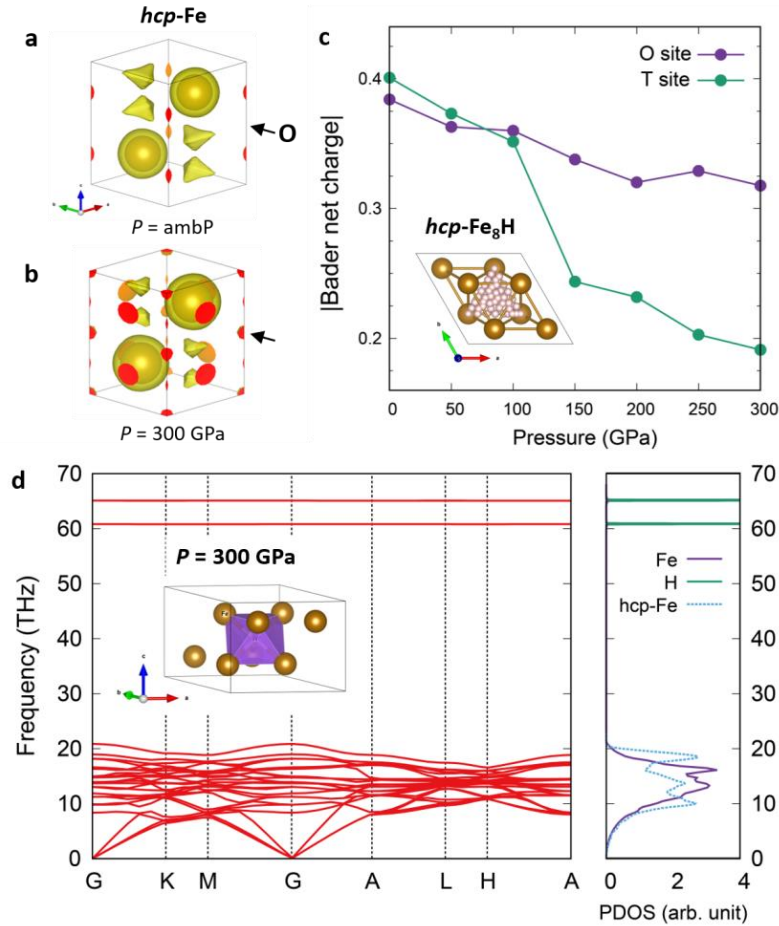

**Fig. 1. a-b.** Electron localization function (ELF) isosurface of pure *hcp*-Fe at ambient pressure and $P = 300$ GPa, respectively (with isosurface cutoff of 0.35). **c.** Absolute Bader net charge of H ions in *hcp*-Fe$_8$H as a function of pressure. Purple and green circles represent the hydrogen anion doped at octahedral (H$^O$) and tetrahedral (H$^T$) sites.

(The inset visualizes all the positions of calculated H ions with Fe-H distance larger than 1 Å.) **d.** Phonon band structures and partial density of states (solid lines) of $Fe_8H^O$ at 300 GPa. The cyan dashed curve is the total density of states of pure *hcp*-Fe at 300 GPa as a comparison.

We start by looking at how the NNA appears in *hcp*-Fe at high pressure conditions and how it offers a pathway toward the superionic state of *hcp*-$Fe_8H$. For *hcp*-Fe, the localizability of interstitial electrons changes as pressure increases, as shown in Figs. 1a and 1b. At ambient pressure, there is no electron localization function (ELF) blob at the octahedral (O) sites – (1/2, 1/2, 1/2), but it clearly appears at 300 GPa, as marked by arrows. (ELF contour plot is shown in Fig. S1). This ELF blob implies the formation of NNAs, the key signature of an electride. In the case of iron *d*-electrons, however, the charge density accumulation is not significant possibly due to the stabilization of *d* electrons at high pressure conditions against the NNA (7). When hydrogen atoms are doped into *hcp*-Fe, nevertheless, the responses of the O sites with the NNA orbital and those of the other interstitial sites are significantly different. Since NNA can interact with and provide electrons to a hydrogen atom (22-25), the formation of NNA can be also chased by observing the Bader net charge of a hydrogen atom as pressure changes.

The Bader net charge of hydrogen atoms doped at all possible interstitial positions with the Fe-H distance larger than 1 Å is investigated. (All interstitial positions are described in the inset of Fig. 1c.) The full results are shown in Fig. S2, and the results of octahedral (O) and tetrahedral (T) – (2/3, 1/3, 5/8) – sites as the most representative cases are shown in Fig. 1c. Near ambient pressure, the Bader net charge of $H^O$ and $H^T$ – hydrogen atom doped at the O and T sites, respectively – does not differ much, and the $H^T$ has slightly larger charge possibly due to the shorter Fe-H bond length. However, $H^O$ begins to possess a larger charge than $H^T$, as the pressure exceeds 100 GPa. The decreasing slope also significantly differs, which means that the Bader net charge of $H^O$ is more tolerant of pressure change than that of the H ions in the tetrahedral or the other sites, where the electrons escape rapidly under pressure. This remarkable difference between $H^O$ and $H^T$ is due to the formation of the NNA at the O site at high-pressure conditions, where the transition point of NNA formation can be represented with the crossing point of Bader net charges at around 100 GPa. The decreasing trend of the Bader net charges can be understood by the competition between the iron *d*-orbitals and *s*-like NNA orbitals. Since the level of *d*-orbitals is less sensitive to pressure than that of *s*- or *p*-orbitals, electrons stay more likely in the iron *d*-orbitals as pressure increases. (7)

In addition, for all H positions tested, the Bader net charge is predominantly correlated to the distance from the O site. As shown in Fig. S2, the Bader charge decreases rapidly as the H ion moves away from the O site, especially when the H ions are at z = 0.5 plane. Such correlation gets stronger as pressure increases. At ambient pressure, some points can have a similar Bader net charge as $H^O$ or an even higher charge, on the other hand, at 300 GPa, all points have a smaller charge than $H^O$. These indicate that *hcp*-Fe (and *hcp*-Fe$_8$H) shows the signatures of the electride state with NNAs formed at the O site at high pressure conditions. As in previous works on electride (12, 18, 30), we will use the term for the electride state as $Fe^{\delta+}(\delta e^-)$, which also can be abbreviated as $Fe:e^-$ for the sake of simplicity. Here, $\delta$ indicates the slight charge accumulation at the O site. For the hydrogen-doped case, we will also use $FeH_x:e^-$ to notate both the sub-stoichiometric hydrogen and the remaining NNAs at vacant O sites. We also note that the dominance of the Bader net charge at the O site was observed even for the case with higher H content, e.g., Fe$_4$H.

The phonon and electronic structures of $Fe_8H^O:e^-$ can elucidate the relationship between the electride and the superionicity. With the NNA formed at 300 GPa, the doped H anion displays flat phonon bands for all translational modes over the whole Brillouin zone, as shown in Fig. 1d. These highly localized vibrational modes indicate that the H atom is under extremely weak interaction potential with *non-bonding* characteristic, i.e., behaves like a rattler (31,32). As shown in the partial phonon density of states (PDOS) in Fig. 1e, these flat modes do not contain Fe contribution, and Fe PDOS also does not have any contribution from H atoms. The independence of the phonon modes between neighboring Fe and H atoms is peculiar, and it implies that the H anion can freely move with shallow harmonic potential well of the Fe: $e^-$ lattice. The stable iron lattice maintained by the presence of NNA while the mobile H ions freely come in and out of the NNA sites is nothing but the superionic state. The flat H-driven phonon bands can then be the prerequisite of the superionic state, and the role of the NNA for the structural stability is necessary. Such role of the NNA was partially discussed and observed for the open structure metal at high pressure conditions (19-20). We also checked that the NNAs were stable regardless of the second doping in other O sites, and the possible coupling between NNA and H was similarly proposed in other works (24).

From the evolution of the electronic structures under pressure, we also found that the H anion reflects the characteristics of the NNA so that becomes electronically separated from the Fe atoms. As shown in Fig. S3, the bandwidth of H 1s states does not increase as pressure increases from ambient pressure to 300 GPa. Rather it decreases by about 50 meV (-4%) while the bandwidth of Fe states increases by about 4 eV (+70%). At ambient pressure, there is a finite H contribution to the Fe bands around -7 eV with a sharp peak structure due to the flatness around the Γ point. Interestingly, this hybridization behavior completely disappears at 300 GPa, which is unusual considering the extreme pressure condition. We expect that it is deeply related to the localized characteristic of the NNA and its coupling to H ions, which reinforces the non-interacting behavior between H and Fe ions. This can also play an important role in maintaining the stable $FeH_x: e^-$ lattice. In this picture of the possible entanglement between the electride and the superionic state, the NNA directly interacts with the H atom, providing a localizable electron at high pressure conditions. Hence, a direct Fe-H bond or the change of a Fe-Fe bond is unnecessary for becoming a charged H anion. As a result, the harmonic potential of the *hcp* iron lattice strengthened by the NNA remains independent of the doped H anion and stable under its diffusion.

## II. Evolution of superionic state under pressure

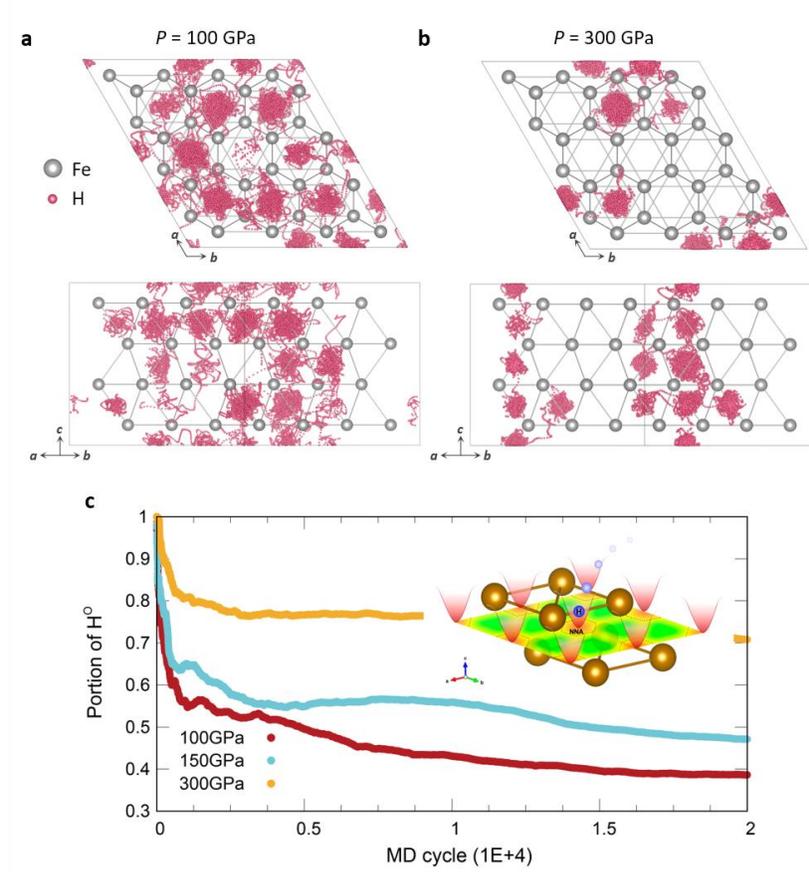

**Fig. 2 a-b.** Molecular dynamics (MD) trajectory of one H anion for superionic *hcp*-FeH$_{0.25}$ state at $P$ = 100 and 300 GPa with the isothermal condition, respectively. ($T$ = 3000 K) Here, we draw Fe atoms as fixed reference points since they harmonically vibrate. **c.** The portion of H anions ($p(H^O)$) near the octahedral sites as a function of MD cycles at $P$ = 100, 150, and 300 GPa. The criteria are set as 0.3·(Fe-H bond length at each pressure condition) of the distance between the H anion and the position of the O site (0.5, 0.5, 0.5). Inset describes the diffusion process of H anion around the NNA site schematically. (2D display of ELF indicates the NNA formation at the octahedral sites marked with dashed hexagons.)

Next, the pressure evolution of the superionic state at isothermal conditions was examined using the *ab initio* molecular dynamics (AIMD) simulation to be compared with the electride behavior. 4×4×2 supercell structure of *hcp*-FeH$_{0.25}$ was investigated at $T$ = 3000 K and $P$ = 15, 100, 150, and 300 GPa conditions. As shown in Figs. 2a and 2b, H anions are more captured around the O sites as pressure increases, consistent with the formation of the NNA at the O sites. At 100 GPa, H anions are distributed relatively sparsely throughout the whole cell. On the other hand, at 300 GPa, H anions are observed mainly around the O sites. (There are also shown the anisotropy of diffusion trajectory (33), and we will discuss it in detail later.) To quantify this difference, the portion of H anions near the O sites (i.e., $p(H^O) = n(H^O)/n(H^{tot})$) as a function of the number of cycles is plotted in Fig. 2c. Here, the H anion near the O sites was defined as the ion with the distance from O sites smaller than 0.3 times of neighboring Fe-H distance. The $p(H^O)$ rapidly decreases at the beginning as H anions quickly diffuse into the cell, but soon it converges to a constant. As the pressure increases, the converged $p(H^O)$ value increases significantly – 38.61%, 47.11%, and 70.80% for 100 GPa, 150 GPa, and 300 GPa, respectively. A different distance cutoff was also verified, for example, 0.5 times of Fe-H distance, which shows the same trend.

It indicates how much O sites are thermodynamically favored. In other words, the H anion at the NNA site is energetically more stable at higher pressure condition. The inset of Fig. 2c schematically described the motion of H anions under harmonic potential well near the O site. The NNAs are formed at O sites of the *hcp* structure at above 100 GPa, and they offer an attractive potential well for the H anions so that they mostly reside there. But the potential well is shallow due to the non-bonding characteristics, so that H anions either stay there or easily diffuse to other O sites. Also, adding and removing the H anion does not affect the stability of the *hcp* cell, as will be discussed in the next part. It is worth noting that the low-pressure condition of 15 GPa was also examined, but the *hcp* cell completely collapsed. This pressure condition was in a range where the NNA could not be observed.

**III. Potential energy surfaces evolution under pressure**

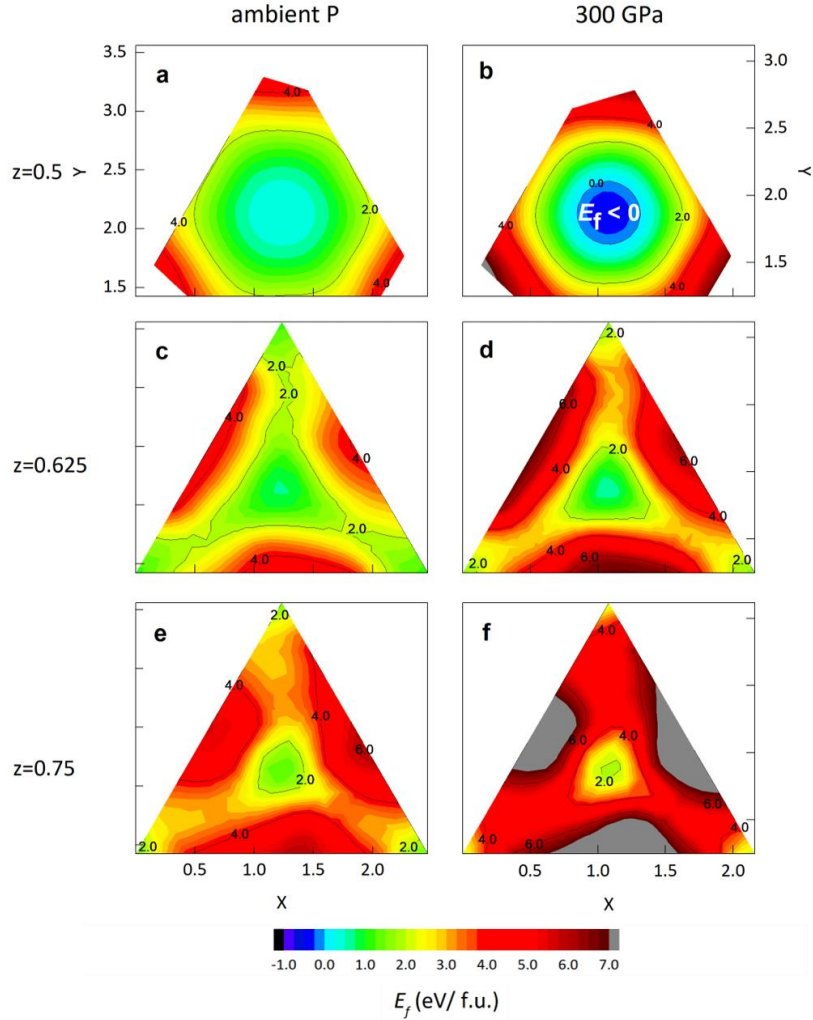

**Fig. 3 a-f.** Potential energy surface for z=0.5, 0.625, and 0.75 planes of *hcp*-$Fe_8H$ for the pressure condition of ambient pressure (left column) and $P$ = 300 GPa (right column). The color bar indicates the formation energy in eV/f.u., and the positions of the calculated H anions are depicted in the inset of Fig. 1c.

The potential energy surfaces (PES) obtained from the DFT calculation elucidate the above MD results. As shown in Fig. 3, the PES for z = 0.5, 0.625, and 0.75 lattice planes were obtained by moving an H atom as a probe in the iron sublattice. The contour plot represents the formation energy referenced by the solid or gaseous hydrogen and iron (34-35). First, we could see that the H ion can be stabilized when it is bound to the NNA. The formation energies at ambient pressure are all positive, explaining the thermodynamic instability of the *hcp* lattice under hydrogen doping.

At 300 GPa, the formation energy for all positions increases further as pressure increases, while only the O site shows the opposite trend to become negative, as marked in Fig. 3b. Contrary to the usual expectation that the densely packed iron atoms can hardly bear H anions at high pressure condition, our result shows the stabilization of H anions at the O site. As observed from the Bader net charge results above, the empty NNA orbitals become active under H doping, giving a higher Bader net charge than other interstitial sites. The coupling between the NNA and H atom seems to stabilize the local charge at the O site so that the formation energy becomes negative.

The PES also elucidates the anisotropic diffusion pathway. There is a smaller energy difference between the in-plane (xy) and out-of-plane diffusion pathways at ambient pressure. But at 300 GPa, the barrier from the O sites to the triangular corner becomes higher to restrict the in-plane diffusion of H anions. It is well captured by looking at the $z = 0.625$ and $z = 0.75$ planes. The potential energy curve for three representative diffusion paths is also shown in Fig. S4. Three paths connect the O site at (0.5, 0.5, 0.5) and the nearest stable O or T sites. The kinetic energy barrier for the path from one O site at $z = 0.5$ plane (O1) to the next O site at $z = 1.0$ plane (O2) is the lowest, and the pressure further increases the relative energy barrier difference compared to the other paths. At 300 GPa, for example, the kinetic barrier of the O1-O2 path is twice as small as that of the second probable O-T path. This is consistent with the anisotropic H diffusion observed in MD simulation at 300 GPa, as the H anion moves around the O site and then predominantly diffuses along the c axis to the next O sites. It rarely diffuses along the *ab* plane, unlike the 100 GPa case.

As indicated from the potential energy surface, the pressure makes the O site doping more stable than other sites. And the energy barrier for the diffusion also increases as pressure increases, so the critical temperature of the superionic transition should increase. This implies that even though the formation of the NNA may help establish the superionic state of iron-light-element alloy, it does not result in *faster* diffusion. This expected phenomenon was also observed in the MD simulation under IC conditions, as the diffusion coefficient decreases at higher pressure conditions for the case of $FeH_{0.25}$ and $FeO_{0.0625}$ at the same temperatures (26). We also note that the formation energy at 360 GPa, which corresponds to the exact IC pressure condition, is about -840 meV/f.u., more than twice that at 300 GPa, which is about -340 meV/f.u..

**IV. Effects of secondary metal or different anionic elements**

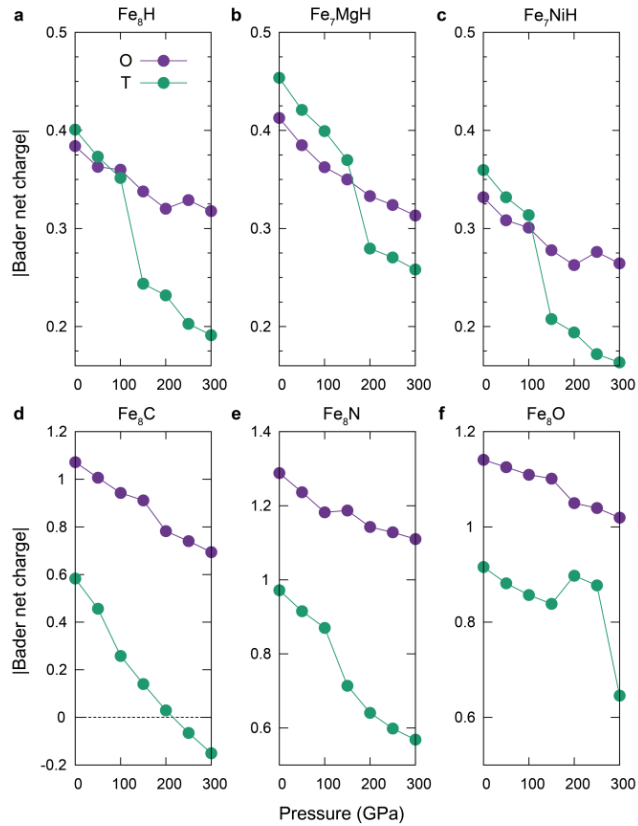

**Fig. 4** Evolution of the Bader net charge as a function of the pressure of **a.** *hcp*-$Fe_8H$, with partial substitution of the secondary metal element Mg or Ni - **b.** $Fe_7MgH$ and **c.** $Fe_7NiH$, and with anion substitution to carbon, nitrogen, or oxygen – **d.** $Fe_8C$, **e.** $Fe_8N$, and **f.** $Fe_8O$, respectively. Purple and green circles are the Bader net charge of the O and T sites, respectively.

To take a more realistic environment of IC into account, the partial substitution of secondary metal elements such as Ni or Mg was investigated (36). As shown in Figs. 4b and 4c, both $Fe_7MgH$ and $Fe_7NiH$ show the same trend of Bader net charge evolution – $H^O$ and $H^T$ have almost the same Bader net charge and decreasing slope at the low pressure regime but a significant difference at the high -pressure regime. However, the extent of the Bader net charge differs, as Mg substitution gives a larger net charge while Ni gives a slightly smaller net charge. This is because *p*-electrons of Mg produce much stronger NNA and a higher possibility of charge accumulation at the O sites (8). On the other hand, the transition point toward the superionic state, *i.e.,* the crossing point of two curves in the graph, also shifts to a higher-pressure value by ~ 50 GPa in the case of Mg substitution. We speculate that Mg atoms easily give electrons to both O and T sites at relatively lower pressure conditions, so two sites compete. In the case of Ni, that transition point does not seem to change.

The effects of other anions, such as carbon, nitrogen, and oxygen, were also investigated. These volatile elements are also possible constituents in IC, and the stable *hcp* structure above 100 GPa and the possible superionic state were reported (26, 37-39). As shown in Figs. 4d-4f, all anions have larger Bader net charge when doped at the O site at ambient pressure, unlike the hydrogen-doped case. Hence, it is hard to see a noticeable crossover of two net charge curves as dramatic as the H case, but the decreasing slopes are still significantly different. As seen from the ELF evolution as depicted in Fig. S5, a larger difference in Bader net charge implies the formation of the NNA at the O site with increasing pressure. Lastly, the evolution of the potential energy curve along the O-T diffusion path under pressure changes is shown in Fig. S6, which is also consistent with the H-doped case– the O site becomes more stabilized, and the barrier energy becomes anisotropic as pressure increases.

**Discussion and Conclusion**

We also note that the effect of the NNA on the doped volatile elements may contribute to solving an important volatile missing problem in geoscience. The existence of volatile elements in deep Earth and the role of the deep reservoir in their cycle are still under debate. For example, a recent study discovered that a significant ratio of nitrogen comes from deep inside the Earth's crust, giving a new hint to the 'nitrogen missing' problem (40). Our NNA-driven approach implies that *hcp*-Fe can act as a reservoir for volatile elements and contribute to their diffusion behavior so that it can be an overarching solution for the missing volatile issues and their cycle.

In this work, the entanglement of the electride and superionic state of iron-light element alloys has been investigated. The NNAs appear at high pressure conditions above 100 GPa, and the H anion doped at that position becomes chemically inert, reflecting the NNA's electronically localized characteristic. This enables the *non-interacting* Fe-H bond, therefore, helps to construct the superionic environment – static Fe and freely moving H ions. The evolution of the superionic state was also investigated using the DFT and MD complementarily, which elucidates the overall diffusion behaviors. Finally, the effect of the secondary metal cations and the other volatile species was verified, and it was shown that the NNA-driven phenomena depend on the major iron.

We suggest that the NNA phenomena and the volatile diffusion are intertwined. It would also give the general implication for rocky planetary cores since this relationship originated from the intrinsic property of the metal. Overall, our study provides an in-depth understanding of the superionic transition under pressure, which provides a new clue about the relationship between the electride and superionic phenomena.

**Materials and Methods**

All our DFT calculations were done by using the projector-augmented wave method (PAW) (41,42) embedded in the Vienna *ab initio* simulation package (VASP) (42-44). In all calculations, the generalized gradient approximation (GGA) of Perdew-Burke-Ernzerhof (PBE) (45) was used for the exchange-correlation functional with the plane-wave cut-off energy of 600 eV. The Monkhorts-Pack *k*-point mesh with $0.03 \cdot (2\pi/\text{Å})$ and $0.02 \cdot (2\pi/\text{Å})$ mesh resolution was used for the unit cell optimization and electronic structure calculation, respectively. At each pressure condition, the unit cell volume was optimized for both *hcp*-Fe and *hcp*-Fe$_8$(H, C, N, or O) with the force convergence criterion of 1E-4 eV/Å. Then the electronic structure and the total energy were calculated with the self-consistent-field energy convergence criterion of 1E-6 eV. The phonon band structure and the density of states were calculated by using the density functional perturbation theory (DFPT) as encoded in VASP to calculate the force constants in real space and PHONOPY. (46)

*Ab-initio* molecular dynamics (AIMD) simulations were conducted to calculate the H ion diffusion behavior in the *hcp*-Fe lattice. VASP was employed for AIMD calculations. We used the PBE exchange-correlation functional and projector augmented wave (PAW) pseudopotentials in the calculations with an energy cutoff of 400 eV. Brillouin zone sampling was performed at the Γ point. For hydrogen bearing structure, 25 atom % hydrogen was randomly placed at the octahedral sites in the *hcp* lattice, and the supercells (4×4×2) for AIMD simulations of FeH$_{0.25}$ contain 80 atoms. The trajectories of H anions were calculated by conducting a grid of NPT ensemble simulations at 3000 K and pressures of 15, 100, 200, and 300 GPa using a Langevin thermostat. The simulations last for 20,000 steps with a step-time of 0.5 fs.

(46) Atsushi T. & Isao T. First principles phonon calculations in materials science. *Scr. Mater.* **108**, 1-5 (2015).**Acknowledgments:** We acknowledge the support of the National Natural Science Foundation of China (42074104, U1930401) and the National Research Foundation of Korea (NRF) (NRF-2020R1A2C1005236, NRF-2020H1D3A2A02111022). Y. H. also acknowledges the support from the Youth Innovation Promotion Association of CAS (2020394).**Author contributions:** I.P., J.H.S., and D.Y.K. conceived the research. I.P. and Y.H. carried out the main calculations. I.P., Y.H., J.H.S., and D.Y.K. analyzed the data and wrote the manuscript. All authors discussed the results and commented on the manuscript.

**Competing interests:** Authors declare that they have no competing interests.

**Data and materials availability:** All data are available in the main text or the supplementary materials.

**Supplementary Materials**

Figs. S1 to S6

# Supplementary Information for

# Origin of superionic state in Earth's inner core

Ina Park[1], Yu He[2,3], Ho-kwang Mao[2], Ji Hoon Shim[1,4,†], and Duck Young Kim[2,‡]

[1] *Department of Chemistry, Pohang University of Science and Technology, Pohang 37673, Republic of Korea*

[2] *Center for High Pressure Science and Technology Advanced Research (HPSTAR), Shanghai 201203, China*

[3] *Key Laboratory of High-Temperature and High-Pressure Study of the Earth's Interior, Institute of Geochemistry, Chinese Academy of Sciences, Guiyang 550081, China*

[4] *Department of Physics, Pohang University of Science and Technology, Pohang 37673, Republic of Korea*

†jhshim@postech.ac.kr

‡duckyoung.kim@hpstar.ac.cn

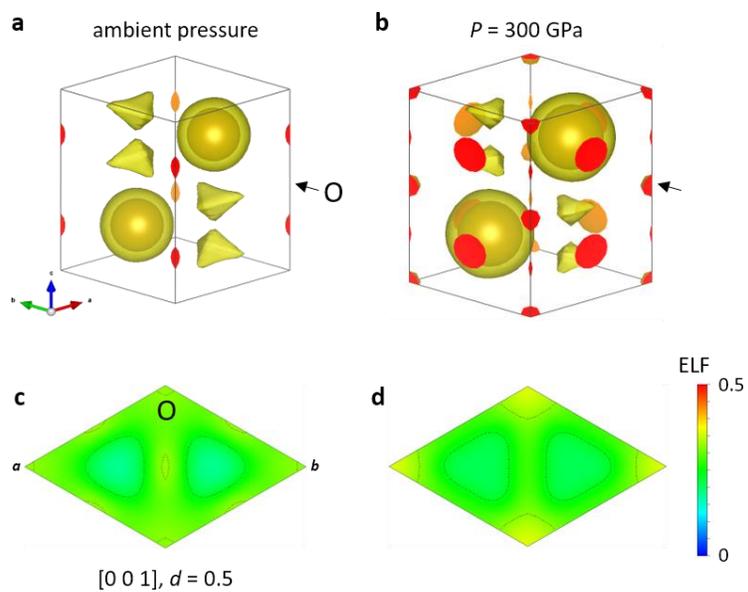

**Figure S1** Electron localization function (ELF) isosurfaces of pure hcp-Fe for **a.** ambient pressure and **b.** 300 GPa. **c,d.** ELF map and contour plot for $z$=0.5 plane for the same pressure conditions.

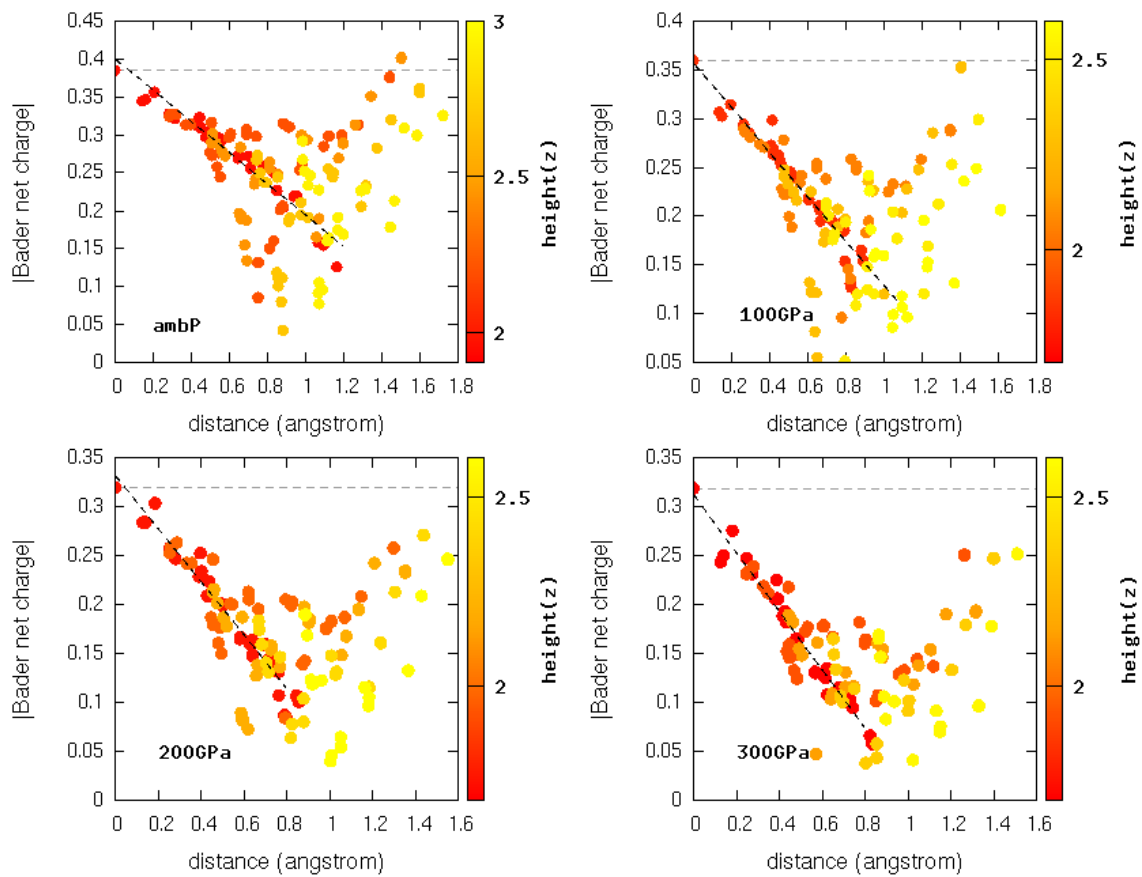

**Figure S2** Absolute Bader net charge of H anions at evenly distributed doping sites as a function of a distance from the O site at each pressure condition. Color of dots represents the z height, where the red dots represent the H anions at the same height of the $H^O$ ($z=0.5$) and yellow dots represent the H anions at $z=0.75$. Black dashed line marks the linear fitted line for the decrease of Bader net charge of H anions at $z=0.5$ plane. Grey horizontal dashed line indicates the value at $H^O$ anion at each pressure condition.

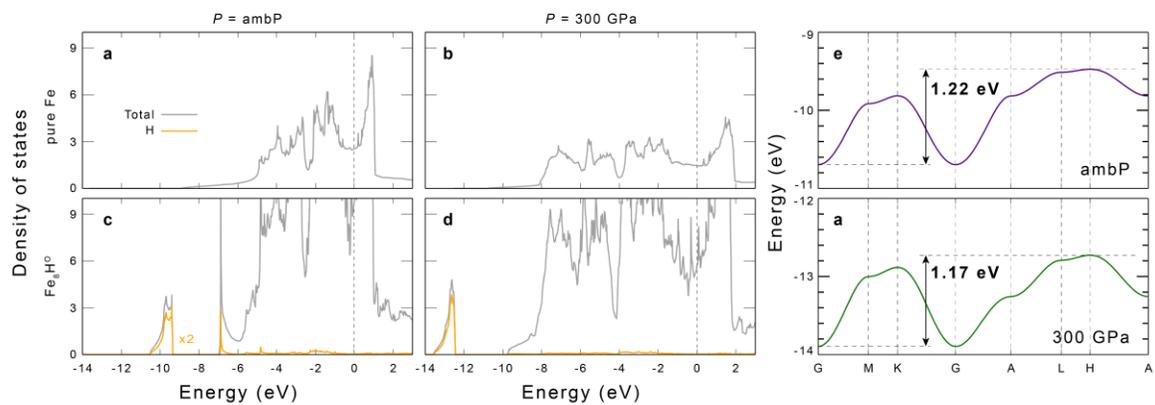

**Figure S3 a-d.** Density of states of pure *hcp*-Fe and $Fe_8H^O$ at ambient pressure and 300 GPa. **e-f.** Band structure of H 1s states at ambient pressure (e) and 300 GPa (f). The bandwidth is indicated with arrow with its value.

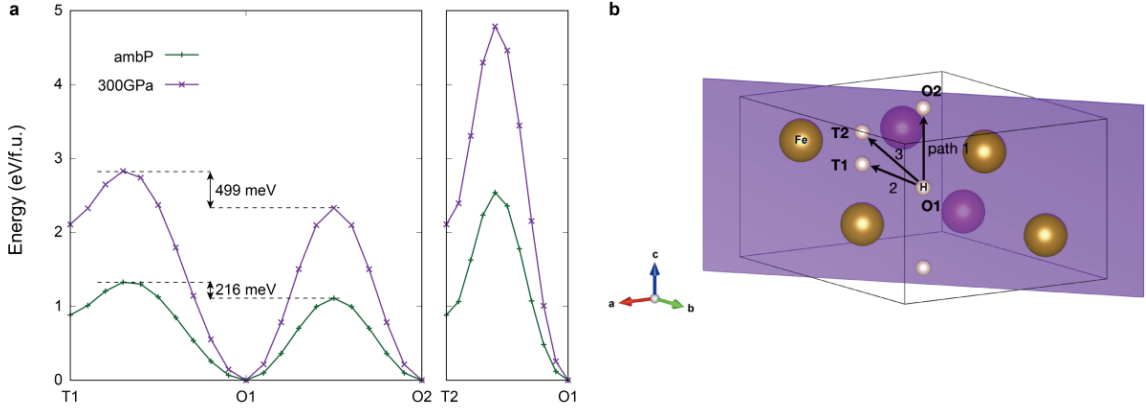

**Figure S4 a.** Potential energy curve for three possible diffusion paths under pressure condition of ambient pressure and 300 GPa. The paths are depicted in **b**, with the background displaying the ELF value of (1 1 0) plane with $d = 1$.

**Figure S5** Electron localization function (ELF) isosurfaces with different anion element cases. **a-b.** ELF of hcp-Fe$_8$C at ambient pressure (ambP) (**a**) and P = 300 GPa (**b**). **c-d.** ELF of hcp-Fe$_8$N at ambient pressure (**c**) and P = 300 GPa (**d**). **e-f.** ELF of hcp-Fe$_8$O at ambient pressure (**e**) and P = 300 GPa (**f**).

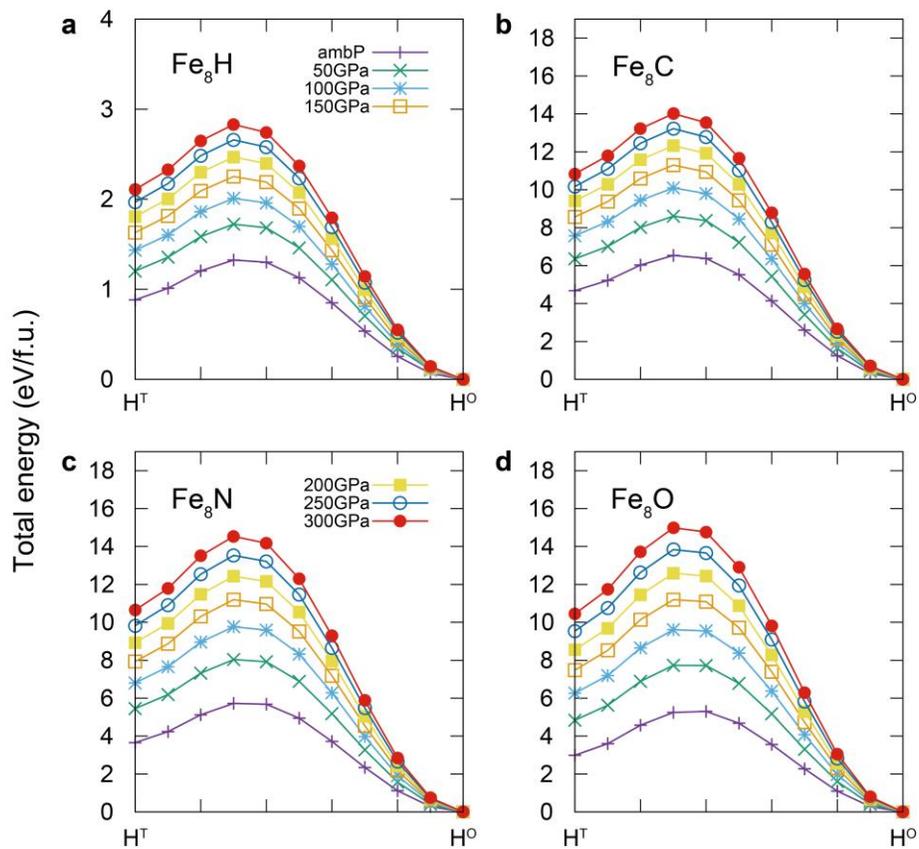

**Figure S6** Potential energy curves for the linear path connecting the O and T site of hcp-Fe$_8$(H, C, N, or O) for the pressure condition from ambient pressure to $P = 300$ GPa.